\documentclass[
amsmath,amssymb,
aps, nofootinbib
rmp, floatfix, reprint,
superscriptaddress,
groupedaddress, showkeys
]{revtex4-2}
\usepackage{graphicx}
\usepackage{dcolumn}
\usepackage{bm}

\usepackage[toc,page,header]{appendix}
\makeatletter
\newcommand\appendix@section[1]{%
  \refstepcounter{section}%
  \orig@section*{Appendix \@Alph\c@section: #1}%
}
\let\orig@section\section
\g@addto@macro\appendix{\let\section\appendix@section}
\makeatother

\usepackage{amsmath,amssymb}
\usepackage{slashed}
\usepackage{graphicx}
\usepackage{dsfont}
\usepackage{bm}
\usepackage[top=1.0in,bottom=1.0in,left=1.0in,right=1.0in]{geometry}
\usepackage[colorlinks,linkcolor=blue,citecolor=blue]{hyperref}
\usepackage{CJKutf8}
\usepackage[caption=false]{subfig}
\usepackage{float}

\begin{document}
\setlength{\oddsidemargin}{0.1cm}
\setlength{\topmargin}{-1cm}
\setlength{\textheight}{22cm}
\setlength{\textwidth}{16cm}
\newcommand{\be}{\begin{equation}}
\newcommand{\ee}{\end{equation}}
\newcommand{\bea}{\begin{eqnarray}}
\newcommand{\eea}{\end{eqnarray}}

\newcommand{\Tr}{\mbox{Tr}\;}
\newcommand{\tr}{\mbox{tr}\;}
\newcommand{\ket}[1]{\left|#1\right\rangle}
\newcommand{\bra}[1]{\left\langle#1\right|}

\newcommand{\avg}[1]{\left\langle #1\right\rangle}
\newcommand{\vnabla}{\mathbf{\nabla}}
\newcommand{\notes}[1]{\fbox{\parbox{\columnwidth}{#1}}}


\title{Dissipative Quantum Dynamics in Static Network with Different Topologies}

\author{Wei-Yang Liu}
\email{wei-yang.liu@stonybrook.edu}
\affiliation{Department of Physics and Astronomy, Stony Brook University, Stony Brook, New York 11794-3800, USA}

\author{Hsuan-Wei Lee}
\email{hsl324@lehigh.edu}
\affiliation{College of Health, Lehigh University, Bethlehem, PA 18015, USA}

\date{\today}


\begin{abstract}

We investigate the dissipative dynamics of quantum population and coherence among different network topologies of a quantum network through the quantum spin model coupled to a thermal bosonic reservoir. Our study proceeds in two parts. First, we analyze a small network of Ising spins embedded in a large dissipative bath, modeled via the Lindblad master equation where temperature arises naturally from system–bath coupling. This approach reveals how network topology shapes quantum dissipative dynamics, providing a basis for controlling quantum coherence through tailored network structures. Second, we propose a mean-field approach extending the scale of the network that captures the dissipative dynamics to large-scale network, connecting network topologies to quantum coherence in complex systems and revealing the sensitivity of quantum coherence to complex network structure. Our results highlight how dissipative quantum dynamics changes in network topologies, providing information for the coherent dynamics of entangled states in network. Our results can be extended to the dynamics in complex systems such as opinion propagation in social models, epidemiology, and various \textcolor{black}{condensed phase} and biological systems.




\end{abstract}

\maketitle

\section{Introduction}

\textcolor{black}{Recent studies have supported that quantum effect plays a crucial role in many dynamical processes of complex systems such as electron transport in a network of quantum dots \cite{greentree2004coherent}, photons in an optical network \cite{schreiber104silberhorn}, the coherent exciton energy transfers in the Light-Harvesting Complex II (LHC2)~\cite{Marais2018-hp, Rozzi2013-hc, Cao2020-jf, Schrodinger2013-wt, Haga2023-kc}, spin entanglement in biological magnetoreception system~\cite{schulten1978biomagnetic,PhysRevLett.104.220502,HoreMouritsen2016}, and entanglement percolation in a complex quantum network \cite{cuquet2009entanglement,Acin:2007}.} 
\textcolor{black}{As a result, new quantum approaches beyond classical descriptions are now in demand to describe the dynamics in complex system.}

\textcolor{black}{A natural and widely used approach for modeling the dynamics in these complex physical systems is to formulate them on networks. Networks are an abstract representation of relationships between nodes, which simplify the complex system as a countable set of nodes, with 
their interactions represented by the edges among nodes. We refer to \cite{BOCCALETTI2006175,Newman_2003} for reviews of network models.}


Classical network models have provided valuable insights into various dynamical processes of complex systems such as
social dynamics and opinion formation 
\cite{castellano2009statistical, acemoglu2011opinion, malik2016transitivity, lu2025agents}, but quantum networks are expected to offer fundamentally different non-local correlations through superposition states and entanglement~\cite{Biamonte:2019, nokkala2024complex,Perseguers:2010}. For instance, 
well-studied classical evolutionary game theory on networks has shown how network topology fundamentally shapes cooperation and collective dynamics \cite{santos2006graph, szabo2007evolutionary, wang2011network} while
quantum game theory has revealed how entanglement between participants fundamentally alters evolutionary stable strategies and Nash equilibria, providing new mechanisms for cooperation and coordination in networked interactions \cite{iqbal2001evolutionarily, zabaleta2017quantum}. See~\cite{biamonte2019complex} (and references therein) for more reviews. 

\textcolor{black}{As a result, the intersection of quantum science and complex network science has opened new avenues for understanding collective phenomena in complex systems that classical network descriptions 
cannot capture. This motivates us to explore how quantum effects influence the interplay between emergent collective behavior and underlying network structure.} 
\textcolor{black}{In this paper, Ising spin models on networks serve as a minimal yet powerful analogy for complex systems, providing a simple platform for studying the interplay between network structure and collective dynamics in complex systems. Building on this framework, we focus on long-lived quantum coherence and coherent quantum dynamics, and investigate the open quantum dynamics of an Ising spin system with arbitrary network topology coupled to a thermal reservoir. We analyze population dynamics, spin–spin correlations, and quantum decoherence across different network topologies, highlighting how quantum effects shape collective behavior in networked spin systems.}

As the full computation of large scale quantum network is resource-demanding, thereby the mean-field approach is introduced in this paper to estimate the quantum dissipative dynamics for a large scale network as a conceptual presentation to shed the light on how the topological features of networks shapes the dissipative quantum dynamics of a complex system.
Future work will extend this framework to larger networks and systematically explore how different network motifs influence quantum coherence.







The paper is organized as follows. In Sec. \ref{sec:II}, we introduce the model Hamiltonian of the quantum Ising network coupled to a bosonic heat bath and present the resulting dissipative quantum dynamics yielded by the Lindblad-based quantum jump simulation. The effect of network topologies on the open quantum dynamics of Ising network spin system are discussed as well. In Sec. \ref{sec:III}, we approximate the
framework to a mean-field based approach as a resolution to the resource-demanding computation for larger number of nodes and explain its validity. 
Our conclusions are in Sec. \ref{sec:V}. Two Appendices are included to complement some of the detailed derivations.

\section{Ising Model on a static network with a bath}
\label{sec:II}
In this paper, we investigate a system composed of Ising spins distributed over a static network with $N$ nodes and $L$ edges, coupled to a bosonic bath. With $N$ nodes given, the Ising network can have $2^N$ quantum states, denoted by a string of binary spin values $\sigma$
\begin{equation}
    |\sigma\rangle=|\sigma_1\cdots\sigma_N\rangle
\end{equation}
where each on-site spin value $\sigma_i$ is assigned to $\pm1$. 

\begin{figure}
    \centering
    \includegraphics[width=1\linewidth]{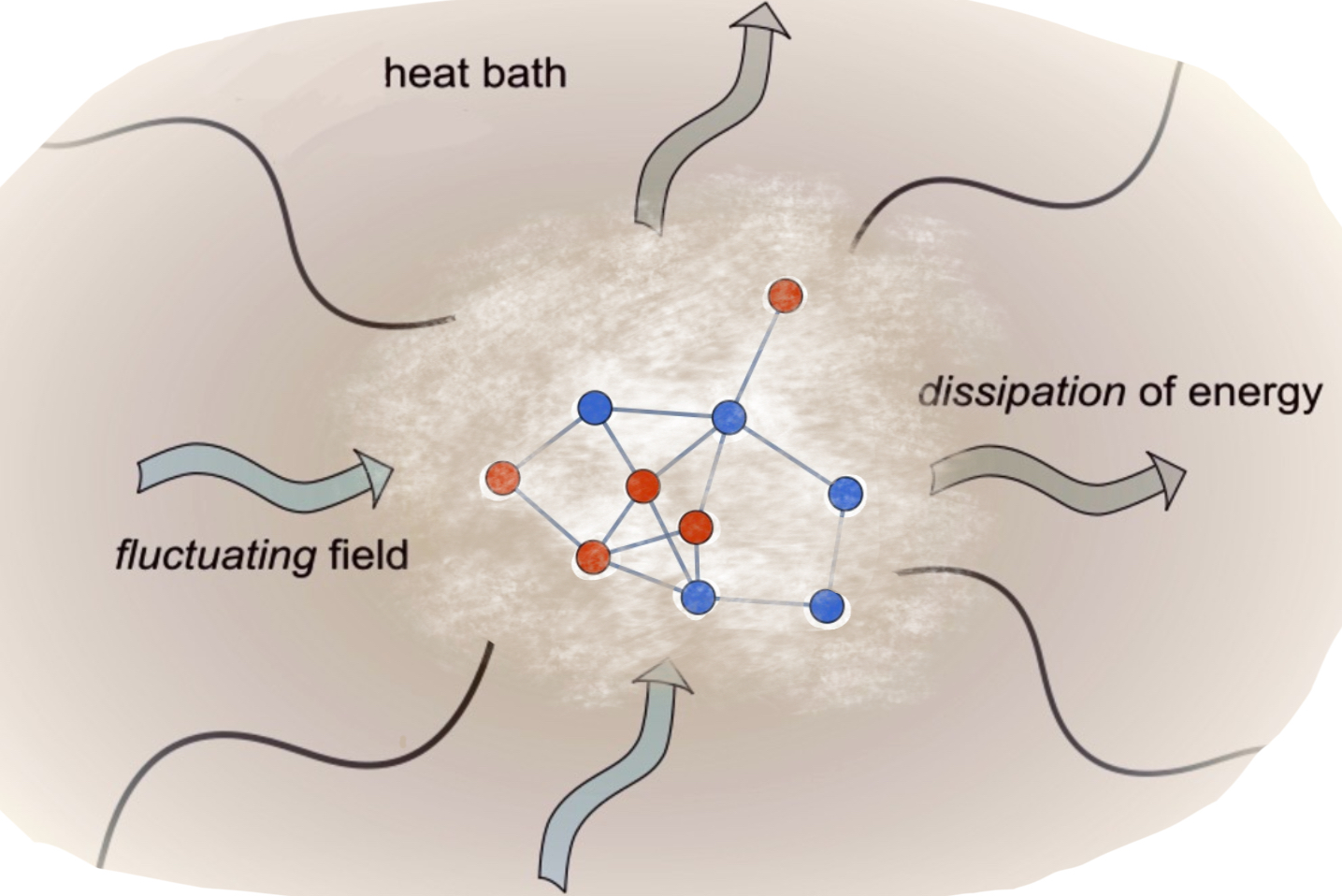}
    \caption{A Ising spin network immersed in the dissipative bath. }
    \label{fig:network-bath}
\end{figure}

As illustrated in Fig.~\ref{fig:network-bath}, in a classical setting, the system interacting with a bath leads to dissipation (friction) , allowing the system's energy to be irreversibly transferred to the environment. This necessarily introduces fluctuating forces acting on the system, and the interplay between dissipation and fluctuations drives the system toward thermal equilibrium at the temperature given by the bath. However, in quantum dynamics, an additional feature emerges: the system can exhibit coherent behavior, which will be suppressed by the coupling between the system and the environment, a process known as decoherence or dephasing. During decoherence, the information is lost to the environment and the system becomes maximally entangled to its bath. Therefore, in this paper, we aim to uncover the topological mechanism on how these network topologies control the quantum coherence of the system. More specifically, we aim to understand how different node number, average degree, and degree disparity achieve a more long-lived the quantum coherence over time.

For a fixed network topology, average degree and degree disparity are the two key structural characteristics that measure the network topology. The average degree of the network, or network density, is defined by

\begin{equation}
    \bar{k}=\frac{2L}{N}
\end{equation}
which quantifies the average number of edges per node. 
The degree disparity $\overline{\Delta k^2}$ is defined as a measure on the variance of the degree distribution, capturing the extent of inhomogeneity in node connectivity under a fixed network density.
 
\subsection{Spin-bath model}

For a more quantitative description of the open system dynamics of the network, we model the open quantum system by the spin-bath framework, where the system composed of relevant degrees of freedom interacting with a bosonic heat bath described by a collection of quantum harmonic oscillators. 

Generally, the full Hamiltonian can be expressed as the sum of three components
\begin{equation}
H = H_{s}+H_{b}+V
\end{equation}
where $H_{s}$ is the system Hamiltonian, $H_{b}$ represents the bath (environment) Hamiltonian, and $V$ describes the interaction between the system and the bath.  

This framework can apply to many systems including the well-known paradigmatic complex system, Sachdev–Ye–Kitaev (SYK) model~\cite{Morsch_2025}. In this work, we begin with a network of Ising spins coupled to a bosonic heat bath. For an Ising network, the system Hamiltonian takes the form:

\begin{equation}
\begin{aligned}
    H_s=&-\sum_{(i,j)\in\rm edges}J_{ij}\sigma^z_i\sigma^z_j-\sum_{i\in \mathrm{nodes}}h_i\sigma^z_i
\end{aligned}
\end{equation}
where $J_{ij}$ is the coupling strength between spins connected by edges in the network. A weak, stationary external field $h_i$ can be applied on each spin to lift the ground state degeneracy of the Ising system. For simplicity, we will only consider the uniform coupling $J_{ij}=J$ and $h_i=h$. The system Hamiltonian $H_{s}$ can be diagonalized in eigenbasis $|\sigma\rangle$ where each configuration $\sigma$ corresponds to an eigenstate with energy $E(\sigma)$. 

\begin{equation}
\begin{aligned}
H_s=&\sum_{\sigma}E(\sigma)|\sigma\rangle\langle\sigma|
\end{aligned}
\end{equation}
Lower-energy configurations are statistically more likely to occur, especially at low temperatures, reflecting the system's tendency to minimize energy. The topology of the Ising network is encoded in the $N\times N$ adjacency matrix $A$ where $A_{ij}=1$ if nodes $i$ and $j$ are connected, and $A_{ij}=0$ otherwise. The energy associated with a given spin configuration $\sigma$ can then be expressed as

\begin{equation}
    E(\sigma)=-\frac J2\sum_{i,j}A_{ij}\sigma_i\sigma_j-h\sum_i\sigma_i
\end{equation}
When $h=0$, the energy is invariant under global $Z_2$ symmetry $\sigma_i\rightarrow -\sigma_i$. The the external field $h$ serves as a symmetry breaking parameter of global $Z_2$ symmetry.

The degrees of freedom associated with the surrounding environment are described by the bath Hamiltonian $H_{b}$  where each bath mode is characterized by a frequency  $\omega_k$, leading to the Hamiltonian:
\begin{equation}
H_{b} = \sum_{k} \omega_{k} (b_{k}^{\dagger}b_{k}+\frac{1}{2})
\end{equation}

This bosonic representation is justified by the central limit theorem and linear response theory, which imply that a large number of weakly coupled, linearly interacting modes give rise to Gaussian statistics. 
The interaction between the system and its environment is described by the potential $V$:

\begin{equation}
V=\sum_{n,k}g_{n,k}\omega_k\left(\sigma^+_n+\sigma^-_n\right)(b_k+b^\dagger_k)
\end{equation}

The interaction to the bath induces spin flips at individual nodes, modeled by the hopping operators  $\sigma^{\pm}_n$, which couple to the local bath coordinates $b_k+b^\dagger_k$. When the system is driven out of equilibrium by external perturbations, the coupling to the bath makes the system relax back to equilibrium. This bath-induced spin-flip dynamics mirrors opinion formation processes in social networks, where agents update their states through environmental influence and stochastic fluctuations, driving the system toward equilibrium configurations \cite{acemoglu2011opinion, malik2016transitivity}. The strength of the coupling between node $n$ and bath mode $k$ is characterized by the coefficient $g_{n,k}$. Practically, these couplings can be given by a spectral density (e.g., Ohmic, sub-Ohmic, super-Ohmic), a band over a certain frequency range, defined as

\begin{equation}
    J(\omega)=\sum_kg^2_{k}\omega_k^2\delta(\omega-\omega_k)
\end{equation}

The spectral density represents how the total variance of the random bath exerting force is distributed over frequency. 
For bosonic bath, the spectral density is typically extended to negative frequencies by imposing the relation $J(-\omega)=-J(\omega)$ \textcolor{black}{\cite{Rivas:2012ugu}}. 

\begin{figure}
    \centering
    \includegraphics[width=1\linewidth]{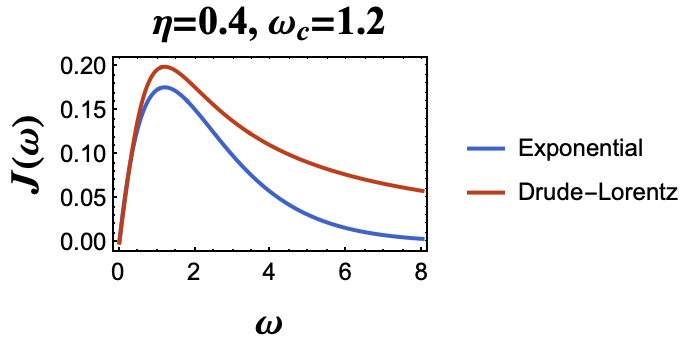}
    \caption{Ohmic spectral density with strength $\eta=0.4$ and cut-off $\omega_c=1.2$ in exponential form (blue) and in Drude-Lorentz form (red)}
    \label{fig:spec}
\end{figure}

In Markovian quantum dynamical systems, the relevant bath correlation functions decay on a time scale $\omega_c^{-1}$ much less than system time scales. This implies that the low-frequency ($\omega\ll\omega_c$) behavior of the spectral density is most important. 
However, in order to avoid certain divergent integrals, a high-frequency cutoff function must be introduced. \textcolor{black}{In Fig.~\ref{fig:spec}, we present two types of the most used spectral density for comparison, the exponential cut-off and Drude-Lorentz cut-off,  with their form defined by}

\begin{equation}
\label{eq:spec_den}
    J(\omega)=\begin{cases}
        \eta \omega e^{\omega/\omega_c} &,\, \mathrm{Exponential}  \\
        \eta \omega \omega_c/(\omega^2+\omega^2_c)&,\, \mathrm{Drude-Lorentz}
    \end{cases}
\end{equation}
where $\eta$ denotes the strength of the spectral density, \textcolor{black}{often referred to as the damping coefficient in the Einstein relation~\cite{bouchaud1990georges,havlin1987diffusion}. In this paper, we model the bath using an Ohmic spectral density ($J(\omega) \sim \omega$) with an exponential high-frequency cutoff. This choice captures linear response of the system and leads to memoryless (Markovian) friction in the system dynamics. The result should not be sensitive to the choice of spectral densities in \eqref{eq:spec_den} as the relevant dynamics is dominated by low frequency region.} 

\begin{figure*}
    \centering
    \includegraphics[width=1\linewidth]{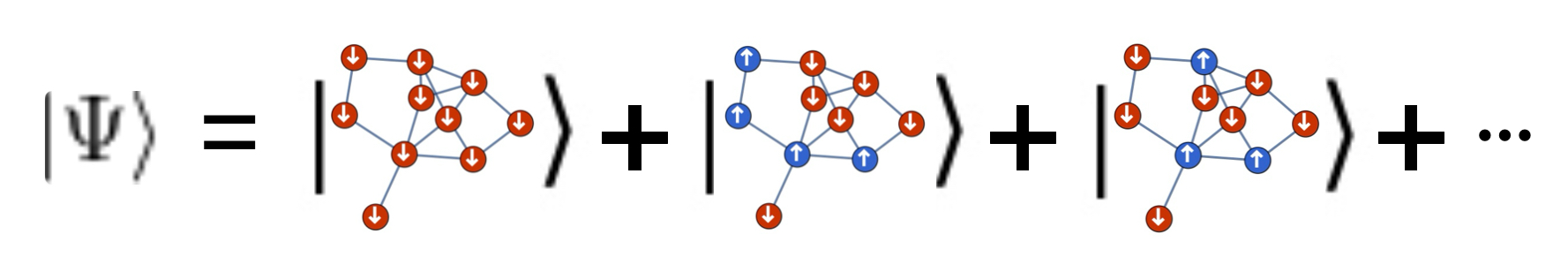}
    \caption{Superpostion of the spin states}
    \label{fig:spin_state}
\end{figure*}

\subsection{Quantum observables of Ising network}

Quantum measurement gives the set of possible outcomes corresponding to physical observables that commute with the system Hamiltonian $H_s$. These observables serve to classify the individual spin configurations within the network. 

One of the simplest observables involves counting the number of spin-up and spin-down states of each node across the network, given by
\begin{equation}
\label{eq:n}
N_{\uparrow,\downarrow}(\sigma)=\sum_{i=1}^N\theta(\pm\sigma_{i})
\end{equation}
where $\theta$ is the unit step function such that $\theta(1)=1$ and $\theta(-1)=0$.

In addition to one-body operators, quantum measurement can also give the spin correlations between connected nodes. Specifically, the number of edges connecting two spin-up nodes, two spin-down nodes, or one spin-up and one spin-down node are defined as:

\begin{equation}
\begin{aligned}
\label{eq:nn}
N_{\uparrow\uparrow}(\sigma)=&\frac12\sum_{i,j}A_{ij}\theta(\sigma_{i})\theta(\sigma_{j}) \\
N_{\downarrow\downarrow}(\sigma)=&\frac12\sum_{i,j}A_{ij}\theta(-\sigma_{i})\theta(-\sigma_{j}) \\
N_{\uparrow\downarrow}(\sigma)=&\sum_{i,j}A_{ij}\theta(\sigma_{i})\theta(-\sigma_{j})
\end{aligned}
\end{equation}

Therefore, each spin state in the network has different degree distribution described by those network operators. In Ising network system, the total number of the spin is fixed, defined as
\begin{equation}
    N=N_\uparrow(\sigma)+N_\downarrow(\sigma)
\end{equation}

We also fix the total number of the spin pair (edge), defined as
\begin{equation}
\label{eq:L}
L=N_{\uparrow\uparrow}(\sigma)+N_{\downarrow\downarrow}(\sigma)+N_{\uparrow\downarrow}(\sigma)
\end{equation} 



With this in mind, the energy of Ising spin network now can be equivalently expressed by the number of nodes and edges.
\begin{equation}
\begin{aligned}
E(\sigma)
=&-J\left(N_{\uparrow\uparrow}+N_{\downarrow\downarrow}-N_{\uparrow\downarrow}\right)-h(N_\uparrow-N_\downarrow)
\end{aligned}
\end{equation}
and the magnetization of the network
\begin{equation}
\begin{aligned}
\label{eq:m}
m(\sigma)
=&N_\uparrow-N_\downarrow
\end{aligned}
\end{equation}
where $\sigma$ denotes the number string of binary spin values on the Ising network.

\subsection{Initial spin states}

Without loss of generality, we assume system starts with a pure state $|\Psi(0)\rangle$ as presented in Fig.~\ref{fig:spin_state}. The initial density matrix can be written as
\begin{equation}
    \rho(0)=|\Psi(0)\rangle\langle\Psi(0)|
\end{equation}

With the bath degrees of freedom traced out, the dissipative dynamics are encoded in the time evolution of coefficient $c_\sigma(t)$. Note that $c_\sigma(t)$ no longer evolve unitarily in time due to the dissipation. The diagonal component (population) describes the probability on each spin configuration and the off-diagonal component (coherence) describes the phase interference between the spin configurations.




\subsection{Quantum master equation on a network}

The reduced density matrix of the system is defined as the total density matrix with the bath degrees of freedom traced out $\rho(t) = \text{tr}_{b}(\rho_{s+b}(t))$. The dynamics between the population and the coherence in the reduced density matrix are not totally independent. As a result, postulating a time evolution operator for the density matrix on phenomenological grounds without rigorous constraints from quantum principles may violate the basic property of density matrix such as positive semi-definiteness~\cite{marquardt2008introductiondissipationdecoherencequantum}. By imposing those constraints on the density matrix such as complete positivity and trace preservation, the evolution of an open quantum system can be expressed through a Kraus decomposition. In Markovian approximation, or Weisskopf–Wigner approximation, the system is assumed to lose significant memory of its past and thus the evolution only depends on the present state. These conditions naturally lead to a specific type of the quantum master equation.









\begin{figure*}
    \centering
\subfloat[]{\includegraphics[width=0.5\linewidth]{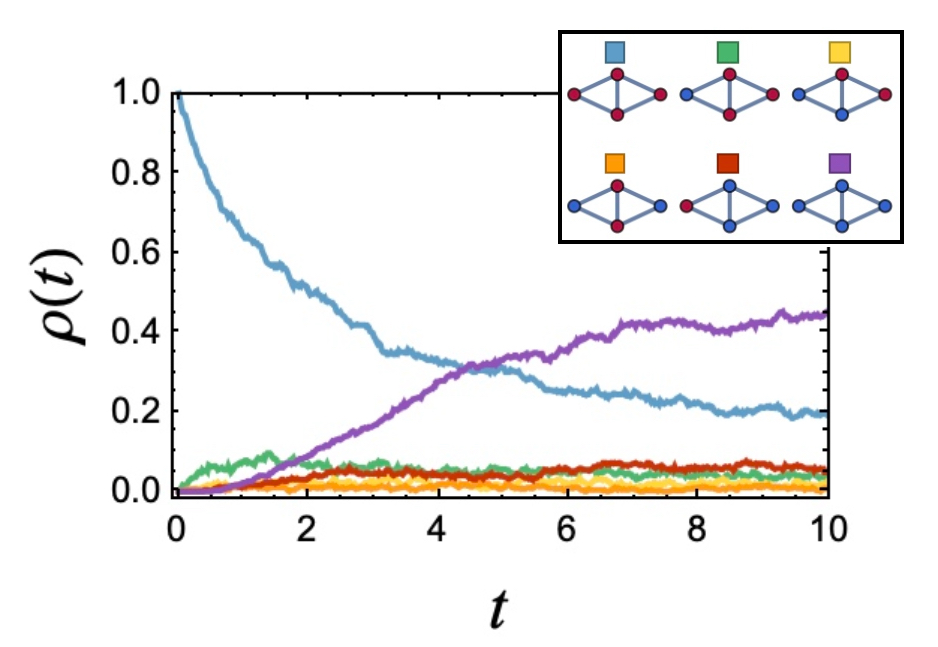}}
\hfill
\subfloat[]{\includegraphics[width=0.5\linewidth]{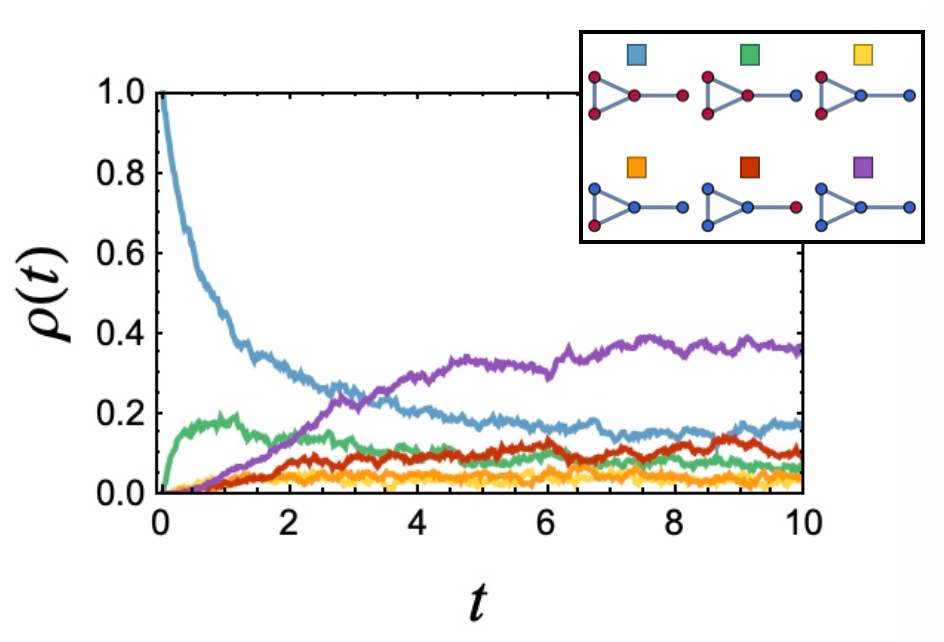}}
\hfill
\subfloat[]{\includegraphics[width=0.5\linewidth]{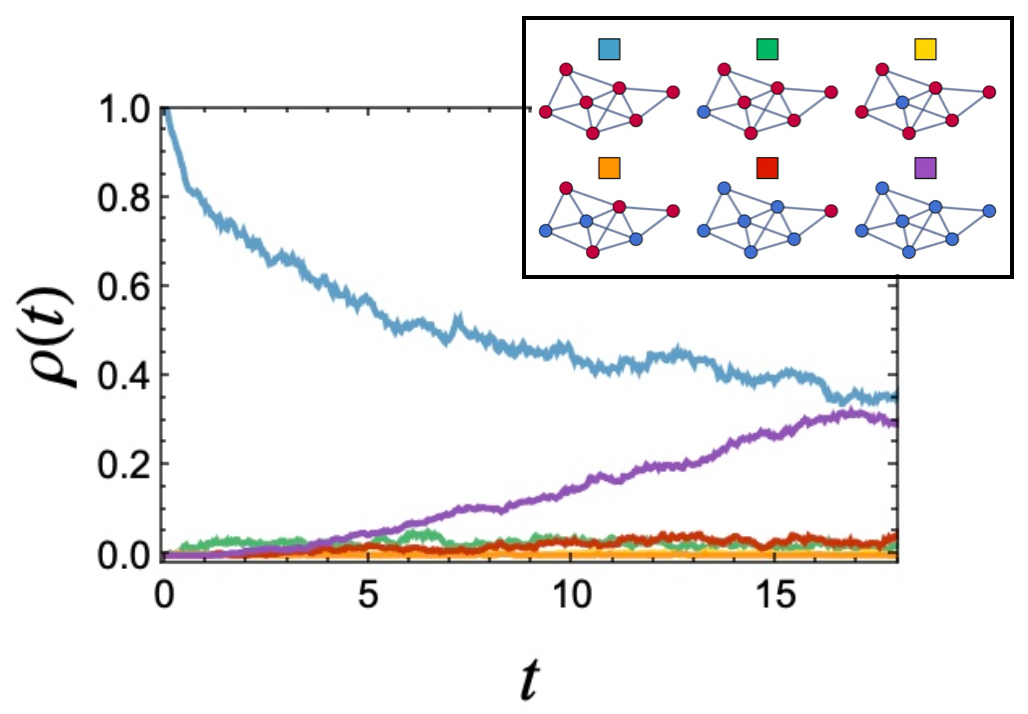}}
\hfill
\subfloat[]{\includegraphics[width=0.5\linewidth]{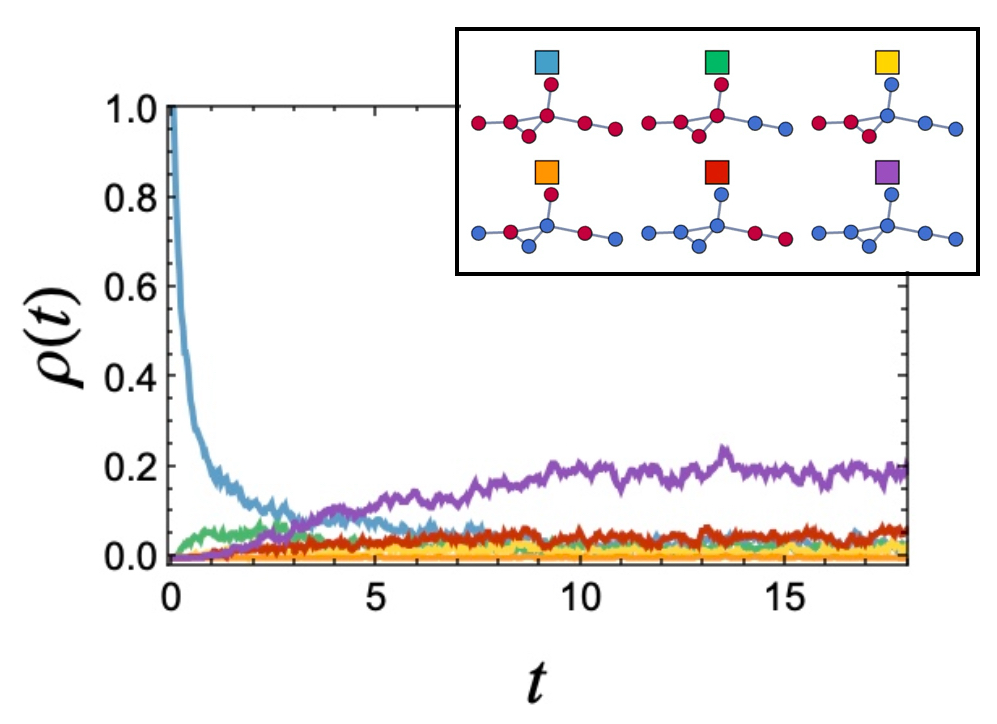}}
    \caption{Time evolution of the diagonal part of the density matrix (population) of 6 different spin states on four different topologies: (a) a network with 4 nodes, 5 edges, and degree disparity $\overline{\Delta k^2}=0.33$; (b) a network with 4 nodes, 4 edges, and degree disparity $\overline{\Delta k^2}=0.67$; (c) a network with 7 nodes, 13 edges, and degree disparity $\overline{\Delta k^2}=1.24$; (d) a network with 7 nodes, 7 edges, and degree disparity $\overline{\Delta k^2}=1.33$. The parameters are fixed by $J=0.4$, $h=0.1$, $\beta=1.2$ with the spectral density fixed by $\eta=0.4$ and $\omega_c=1.2$. The spin states are labeled by colors matching their evolution trajectory and spin values on each node are labeled up (blue) and down (red).}
    \label{fig:quantum_jump}
\end{figure*}

\begin{figure}
    \centering
\subfloat[]{\includegraphics[width=\linewidth]{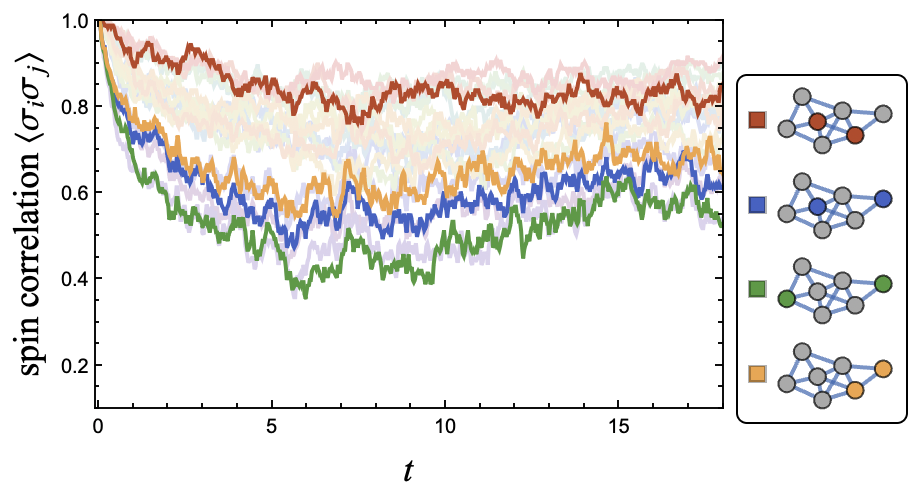}}
\hfill
\subfloat[]{\includegraphics[width=\linewidth]{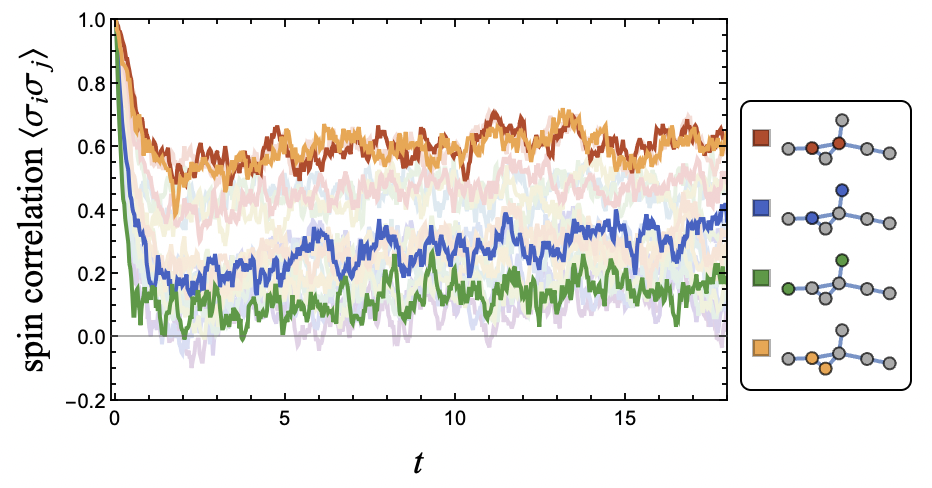}}
    \caption{The spin-spin correlation $\langle\sigma_i\sigma_j\rangle(t)$ between node $i$ and $j$ on the two different 7-node network (a) and (b). Red node: spin down, blue node: spin up. The system parameters is given by $J=0.4$, $h=0.1$, $\beta=1.2$ with spectral density fixed by $\eta=0.4$ and $\omega_c=1.2$.}
    \label{fig:spin-spin}
\end{figure}

If the system-bath interaction is treated perturbatively, time-local Markoian quantum master equation can be written as 

\begin{align}
\dot{\rho}(t) =& -i[H_s,\rho(t)]\nonumber\\
&-\int_{0}^{\infty}d\tau \text{tr}_{b}\{[V,[V(-\tau),\rho(t)\otimes\rho_{b}^{eq}]]\}
\end{align}

This master equation can be recast in Linblad form
\begin{equation}
\begin{aligned}
    &\dot{\rho}(t)=-i[H_s,\rho]\\
    &+\sum_{n}\gamma_n\left(\sigma^+_n\rho \sigma^-_n-\frac12\{\sigma^+_n \sigma_n^-,\rho\}\right)\\
    &+\sum_{n}\gamma_n\left(\sigma^-_n\rho \sigma^+_n-\frac12\{\sigma^-_n \sigma_n^+,\rho\}\right)
\end{aligned}
\end{equation}


For simplicity, we renormalize the Lamb shift into $H_s$.
The transition only happens when the Hamming distance $d_{\rm H}$ between two spin states is 1, i.e. $d_{\rm H}(\sigma,\sigma')=1$. The transition rate $\gamma_n$ between the two states can be calculated by the time correlation of the collective bath coordinate, 
\begin{equation}
   \gamma_{n}=\gamma(\omega_n)
\end{equation}
where $\omega_n=E(\sigma)-E(\sigma')$, the energy differences calculated by flipping spin at node $n$.  
The decay rate for node $n$ flip the spin related to bath is defined as
\begin{equation}
    \gamma(\omega)= 
    \begin{cases}
    2\pi J(\omega)\bar{n}(\omega) &,\ \omega>0 \\
    2\pi J(\omega)[\bar{n}(\omega)+1] &,\ \omega<0
    \end{cases}
\end{equation}
where the boson number in the bath is defined 
\begin{equation}
    \bar{n}(\omega)=\frac{1}{e^{\beta\omega}-1}
\end{equation}
satisfying the Bose-Eistein distribution, as a result of detailed balance condition in thermal equilibrium.
\begin{equation}
    \gamma(\omega)=e^{-\beta\omega}\gamma(-\omega)
\end{equation}

\begin{figure}
    \centering
\subfloat[\label{fig:ent1}]{\includegraphics[width=0.935\linewidth]{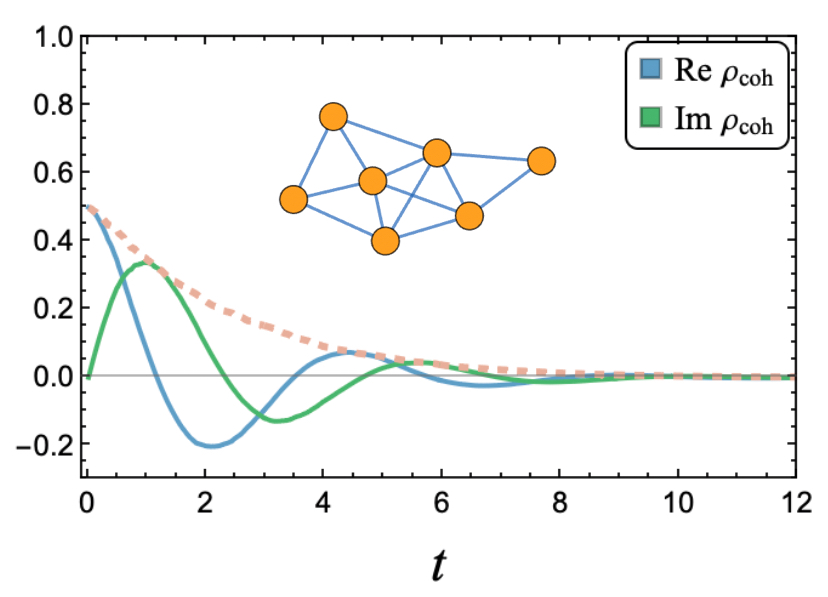}}
\hfill
\subfloat[\label{fig:ent3}]{\includegraphics[width=0.96\linewidth]{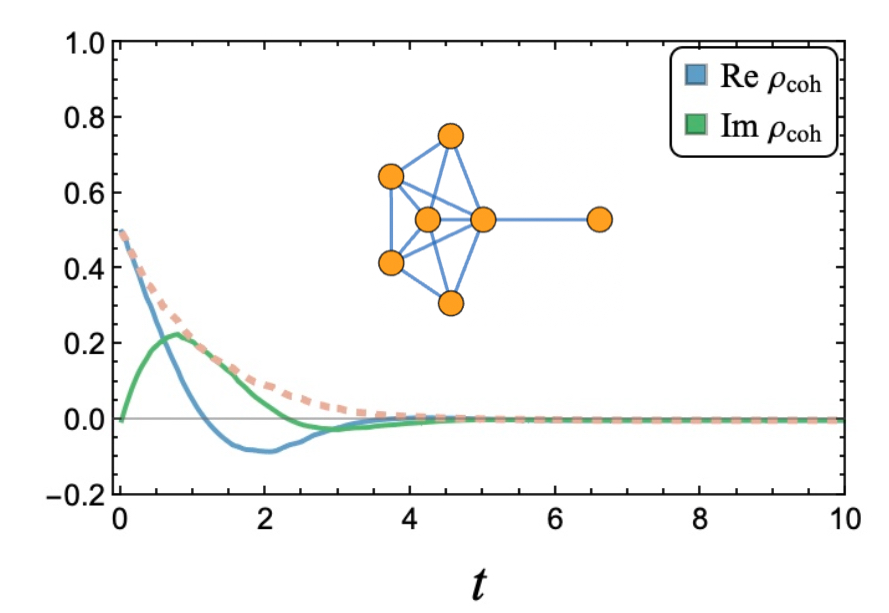}}
\hfill
\subfloat[\label{fig:ent2}]{\includegraphics[width=0.96\linewidth]{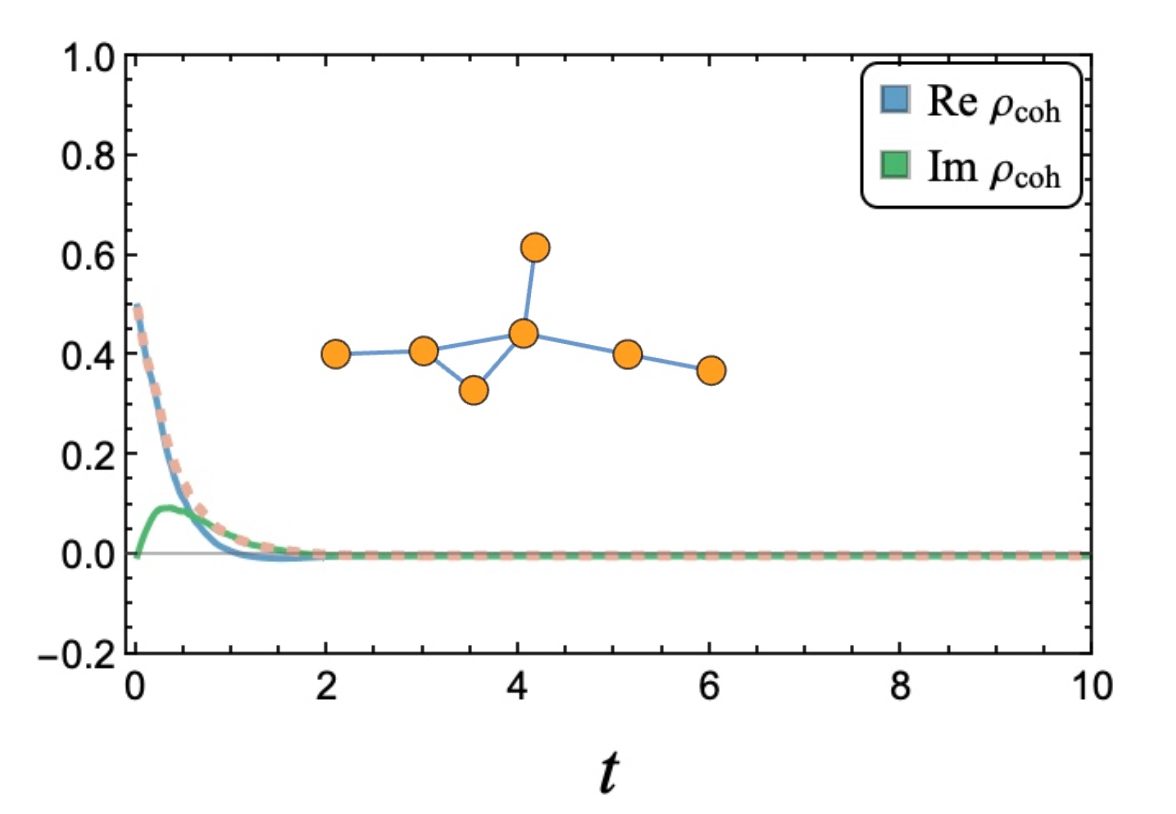}}
\caption{Coherent dynamics on 3 different 7-node networks (a), (b), and (c) with their topologies listed in Table~\ref{tab:tcoh}. The parameters are given by $J=0.4$, $h=0.1$, and $\beta=1.2$, with the spectral density is fixed by $\eta=0.4$ and $\omega_c=1.2$.}
    \label{fig:ent}
\end{figure}

Eventually, the stationary solution should be
\begin{equation}
    \rho_{\sigma\sigma'}(\infty)=\begin{cases}
       e^{-\beta E(\sigma)}/Z &,\ \sigma=\sigma'\\
      0 &,\ \sigma\neq\sigma'\\
    \end{cases}
\end{equation}
with the partition function for the thermal equilibrium ensemble

\begin{equation}
\begin{aligned}
    Z\equiv&\mathrm{Tr}( e^{-\beta H_s})=\sum_{
    \sigma_1\sigma_2\cdots\sigma_N
    }e^{-\beta E(\sigma)}
\end{aligned}
\end{equation}

\subsubsection{Population dynamics}

We start the time evolution with a all-down ground state on different network topologies where the node number ranges from 4 to 7. In Fig.~\ref{fig:quantum_jump}, we present the population dynamics on four of the network configurations. Two of these graphs consist of 4 nodes, one with low and the other with high network density (edge-to-node ratio). The remaining two graphs contain 7 nodes, again for both low and high density cases. 
The results shows that greater node aggregation, or higher node connectivity, leads to slower population decay, indicating a more rigid network. This behavior can be attributed to the fact that a higher average degree (i.e., a larger number of edges) results in a larger energy gap between the ground states (such as all-up or all-down) and the excited states. Consequently, the transition from one ground state to another requires to overcome a higher energy barrier, thereby prolonging the dynamical processes.  Although our results have only been tested on a small network but we expect this is the same case for most common large complex network.

The dynamics of the physical quantity $O$ of the network spin can also be computed by 

\begin{equation}
\langle O\rangle(t)=\sum_{\sigma}\rho_{\sigma\sigma}(t)O(\sigma)
\end{equation}

In Fig.~\ref{fig:spin-spin}, we present the relaxation of the spin-spin correlation on the Ising network with two different 7-node networks, one with low and the other with high network density. This illustrates the transition of the spin-spin correlations on the Ising network from a initial pure quantum state (all-down ground state in this case) to a mixed classical thermal states. In the thermal equilibrium, spin pairs connected by shorter paths exhibit stronger correlations, particularly when they are connected through multiple paths. 
The transition rate is also very sensitive to network density and degree disparity.
A small transition rate in thermalization enables the network to sustain spin–spin correlations for an extended period before reaching thermal equilibrium. The prolonged relaxation in the networks with higher network density and lower degree disparity during the relaxation of these correlations reflects the presence of long-lived quantum coherence.

\subsubsection{Coherent dynamics}

To investigate quantum coherent dynamics in the Ising network, we need to create a collective spin state, i.e. a superpostion of quantum states on the energy degenerate manifold. As presented in Fig.~\ref{fig:spin_state}, the superposition of the initial spin configuration can be written as
\begin{equation}
    |\Psi(0)\rangle=\sum_{
    \sigma_1\sigma_2\cdots\sigma_N
    }c_\sigma(0)|\sigma\rangle
\end{equation}

By interacting with the bath, the dissipation leads decoherence. Once the interference fades out, the probability distribution of the Ising network become classical thermal states.


In this paper, we prepare a entangled Greenberger–Horne–Zeilinger (GHZ) state as the initial state for the time evolution. GHZ states are a special type of maximally entangled quantum states involving three or more nodes (qubits) \cite{Greenberger1990}. \textcolor{black}{They serve as a standard benchmark for multipartite entanglement, analogous to the role played by Bell states in bipartite systems.} In quantum Ising network, it is defined by an equal superposition of two ground states 
\begin{equation}
    |\Psi(0)\rangle=\frac{1}{\sqrt{2}}\left(|\uparrow\uparrow\cdots\uparrow\rangle+|\downarrow\downarrow\cdots\downarrow\rangle\right)
\end{equation}
\textcolor{black}{The discussion regarding the coherent dynamics here are not specific to GHZ states and other entangled initial states are expected to present similar coherent dynamics. Nevertheless, GHZ states are expected to exhibit more pronounced signals of multipartite coherence, making them a useful test case without loss of generality. }

\textcolor{black}{We present the estimation of decoherence time of 5 different networks in Table~\ref{tab:tcoh} where the coherence $\rho_{\rm coh}$ is defined by the off-diagonal component of the density matrix of a GHZ state. The blue curve denotes its real part and the green curve denotes its imaginary part. The decoherence time is obtained by the exponential fit using the function $0.5\exp(-t/t_{\rm decoh})$ for the modulus of coherence $|\rho_{\rm coh}|$ in Fig.~\ref{fig:ent}.} As presented in Table~\ref{tab:tcoh}, the decoherence time evolution is also sensitive to the network topologies. \textcolor{black}{The prolonged decoherence is observed in the networks with higher mean degree $\bar k$ and lower degree disparity $\overline{\Delta k^2}$.} In Fig.~\ref{fig:ent}, we show coherent dynamics of the three 7-node networks from Table~\ref{tab:tcoh} as illustration. This can be used to develop possible network topology that gives the optimal quantum coherence for future development on dissipative quantum dynamics in complex systems.


\begin{table}[]
    \centering
    \begin{tabular}{c|c|c|c|c}
     graph & $N$ & $\bar k$ & $\overline{\Delta k^2}$  &  $t_{\rm decoh}$ \\
     \hline
     
    \includegraphics[scale=0.09]{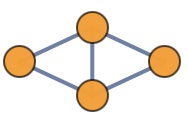}
     & 4 & 2.50 & 0.333  & $1.9083(51)$ \\
     \includegraphics[scale=0.2]{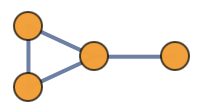}
     & 4 & 2.00 &  0.667 & $0.8600(10)$ \\
     \includegraphics[scale=0.15]{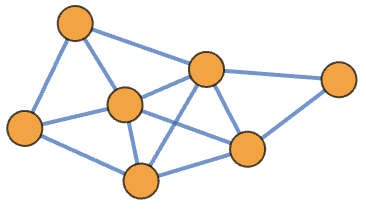}
     & 7 & 3.71 & 1.238  & $2.4377(60)$ \\
     \includegraphics[scale=0.16]{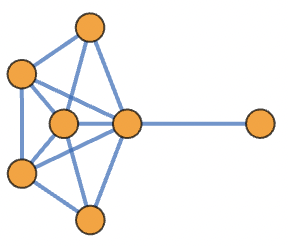}
     & 7 & 3.71 &  2.571 &  $1.1433(19)$\\
     \includegraphics[scale=0.2]{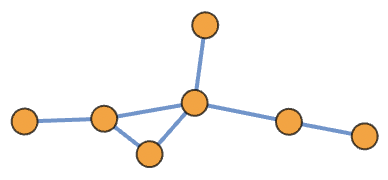}
     & 7 & 2.00 &  1.333 & $0.39824(88)$
    \end{tabular}
    \caption{The decoherence time of the Ising network with the same spin coupling $J=0.4$, external field $h=0.1$, and temperature $\beta=1.2$ where $N$ is total number of nodes, $\bar k$ is network density, $\overline{\Delta k^2}$ is the degree disparity, and $t_{decoh}$ is the decoherence time }
    \label{tab:tcoh}
\end{table}

\section{Mean field approximation}
\label{sec:III}
Extending the computation beyond 10 nodes could become quite computationally demanding. Therefore, we approximate the dynamics using a mean-field approach by assuming the local magnetization of neighbors and the global magnetization are identical.

\subsection{Local magnetization v.s. global magnetization}

The validity of mean field approximation lies in the observation that the local spin average surrounding a node $i$ is the same as the global average. For each spin configuration $\sigma$, the local spin average surrounding the node $i$ is defined by

\begin{equation}
    \bar{s}_i(\sigma)=\frac{1}{k_i}\sum_{j}A_{ij}\sigma_j
\end{equation}
where $k_i$ is the degree of the node $i$ and $A$ is the adjacency matrix of the network. 

\begin{figure}
    \centering
\subfloat[\label{1}]{\includegraphics[width=\linewidth]{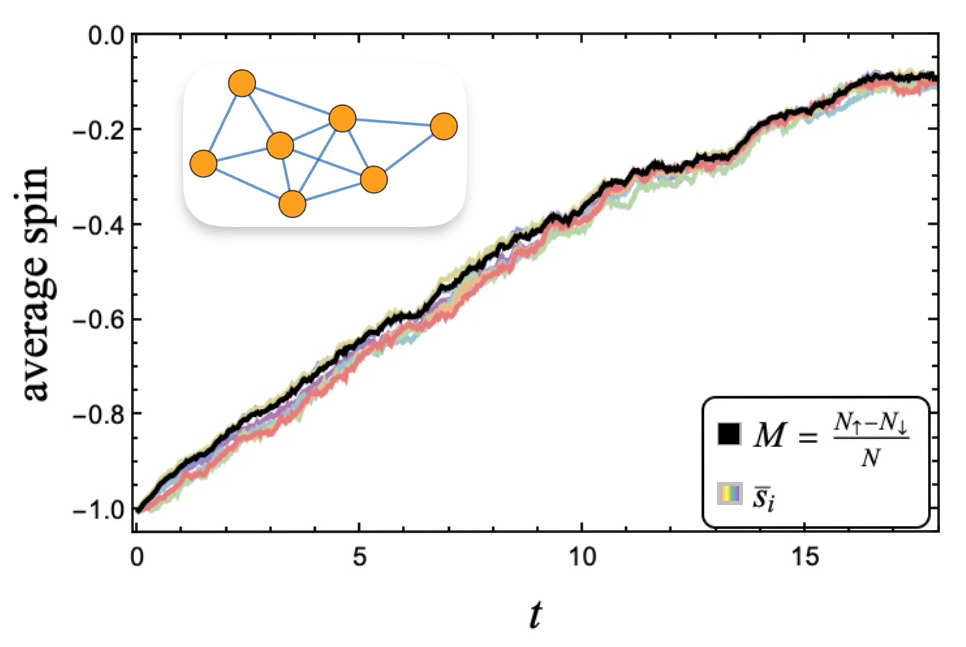}}
    \hfill
\subfloat[\label{2}]{\includegraphics[width=\linewidth]{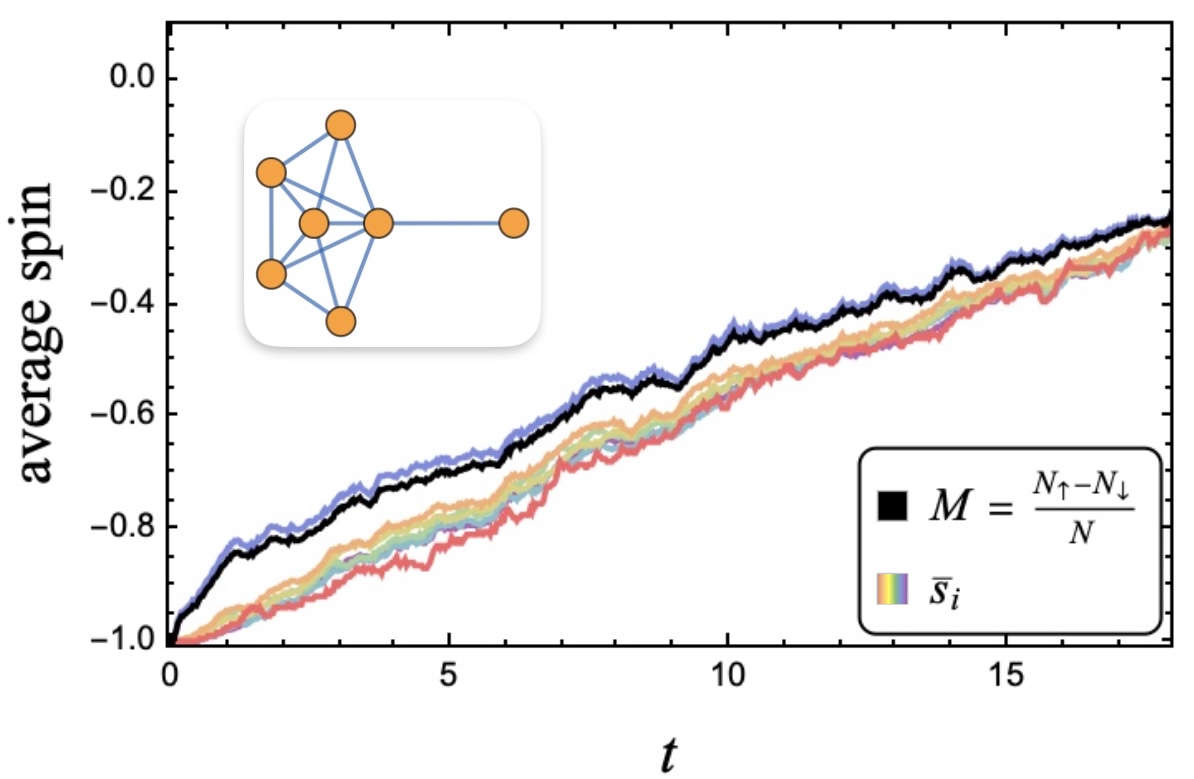}}
    \hfill
\subfloat[\label{3}]{\includegraphics[width=\linewidth]{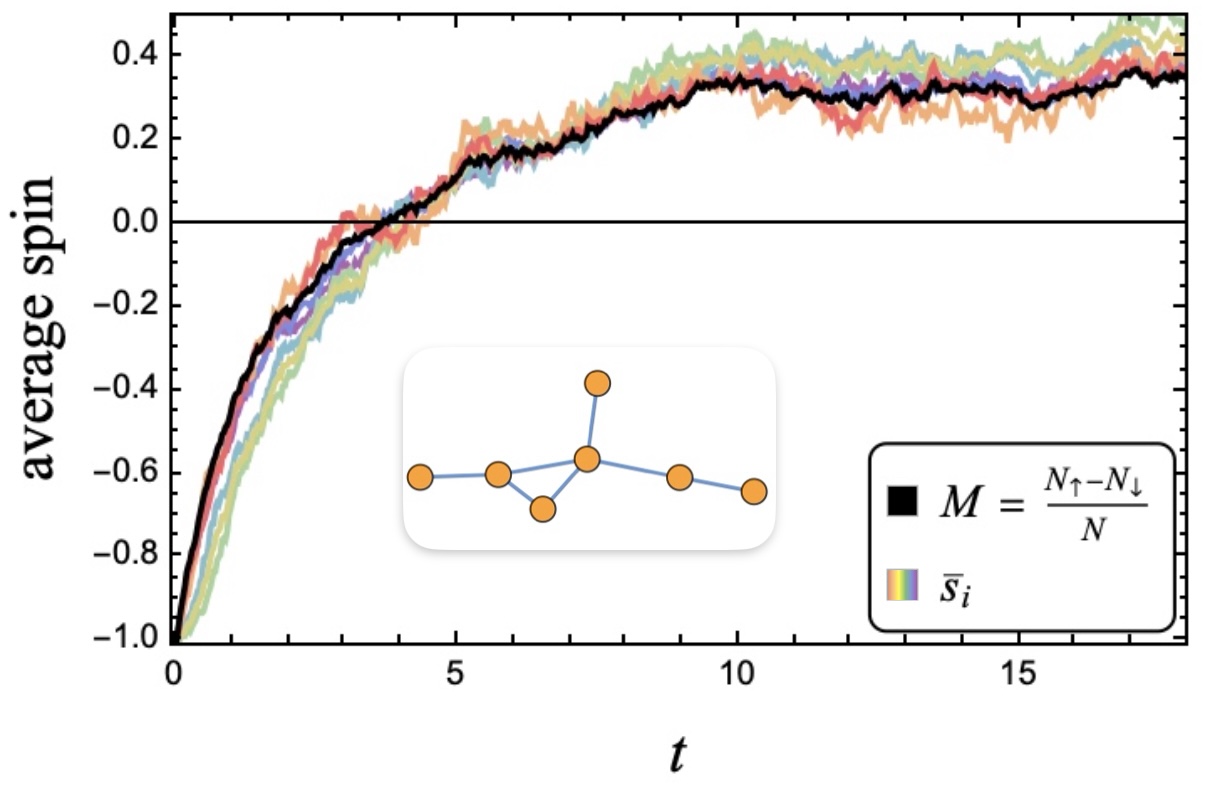}}
    \caption{\textcolor{black}{The comparison of the time evolution between the global magnetization (black curve) and the local magnetization of each node in the 7-node network with different topologies (a), (b), and (c). The trajectories of the 7 nodes are colored sequentially from red to violet in rainbow color scheme, corresponding to nodes 1 through 7.}}
    \label{fig:mean_spin}
\end{figure}

We test the condition on 4-node and 7-node networks. As shown in Fig.~\ref{fig:mean_spin}, the local average spin $\bar{s}_i(t)=\sum_\sigma\bar{s}_i(\sigma)\rho_{\sigma\sigma}(t)$ for each node is consistent with the global magnetization $M(t)=\sum_\sigma m(\sigma)\rho_{\sigma\sigma}(t)/N$. This shows that the mean-field assumption holds,

\begin{equation}
\label{eq:mfc}
    \bar{s}_i(\sigma)\approx\frac{m(\sigma)}{N}\, ,
\end{equation}
indicating the homogeneity of the spin distribution across the network, even for small system sizes.

Given that the network connections are symmetric (i.e., undirected), the mean-field assumption further implies that the local neighborhood of a node is statistically equivalent, regardless of whether the node is in the spin-up or spin-down state. Therefore, the mean degree of the network $\bar{k}$ is equal to the mean degree of the spin-up nodes $\bar{k}_\uparrow$ as well as the spin-down nodes $\bar{k}_\downarrow$.


\begin{equation}
\label{eq:mean_cond}
    z=\bar{k}=\bar{k}_\uparrow=\bar{k}_\downarrow
\end{equation}
where $z$ is the effective dimensionality of the Ising network (coordination number). $\bar{k}_\uparrow$ and $\bar{k}_\downarrow$ are the average degrees of spin-up and spin-down nodes, defined by

\begin{eqnarray}
\label{eq:kud}
\bar{k}_{\uparrow}&=&\frac1{N_{\uparrow}}\sum_ik_i\theta(\sigma_i) \nonumber\\ \bar{k}_{\downarrow}&=&\frac1{N_{\downarrow}}\sum_ik_i\theta(-\sigma_i)
\end{eqnarray}
The details of this derivation are provided in the Appendix~\ref{app:kud}.


With this in mind, now we can reduce the spin configuration dependence into $N_\uparrow$ dependence for many physical observables.
\begin{align}
\label{eq:mf_nn}
    &N_{\uparrow,\downarrow}(\sigma)\rightarrow N_{\uparrow,\downarrow} \nonumber\\[5pt]
&N_{\uparrow\uparrow,\downarrow\downarrow}(\sigma)\rightarrow \frac{z N_{\uparrow,\downarrow}^2}{2N} \nonumber\\[5pt]
&N_{\uparrow\downarrow}(\sigma)\rightarrow \frac{z N_{\uparrow}N_{\downarrow}}{N}
\end{align}
and the energy of a spin state can be expressed as
\begin{equation}
\label{EN}
E(\sigma)\rightarrow E_{N_\uparrow}=-\frac{zJ}{2N}(N_{\uparrow}-N_{\downarrow})^2-h(N_{\uparrow}-N_{\downarrow})
\end{equation}
\textcolor{black}{The details of this derivation are provided in the Appendix~\ref{app:mf}.}

\begin{figure*}
    \centering
\subfloat[]{\includegraphics[width=.5\linewidth]{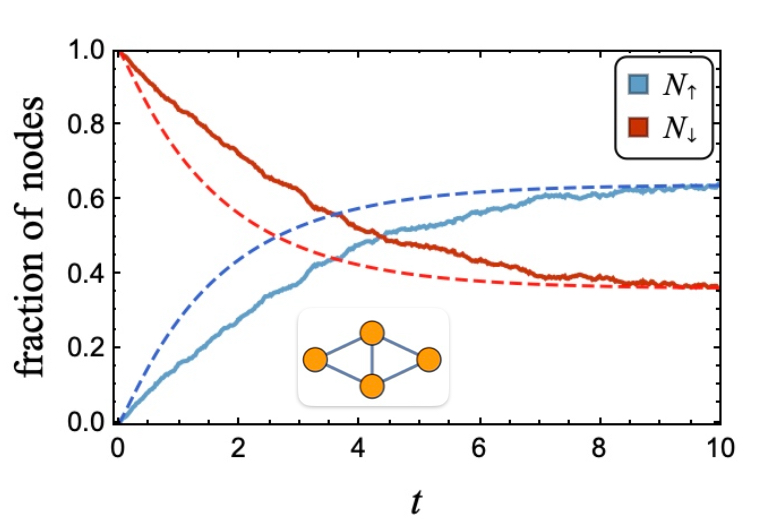}}
\hfill
\subfloat[]{\includegraphics[width=.5\linewidth]{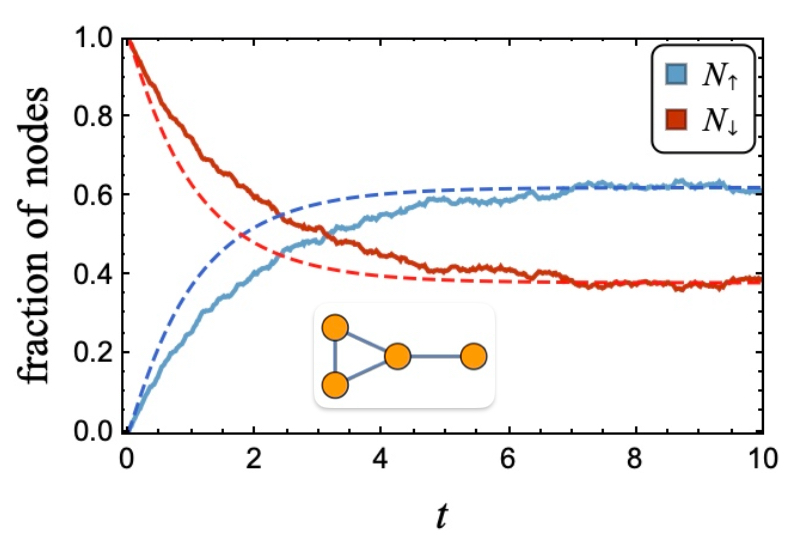}}
\hfill
\subfloat[]{\includegraphics[width=.5\linewidth]{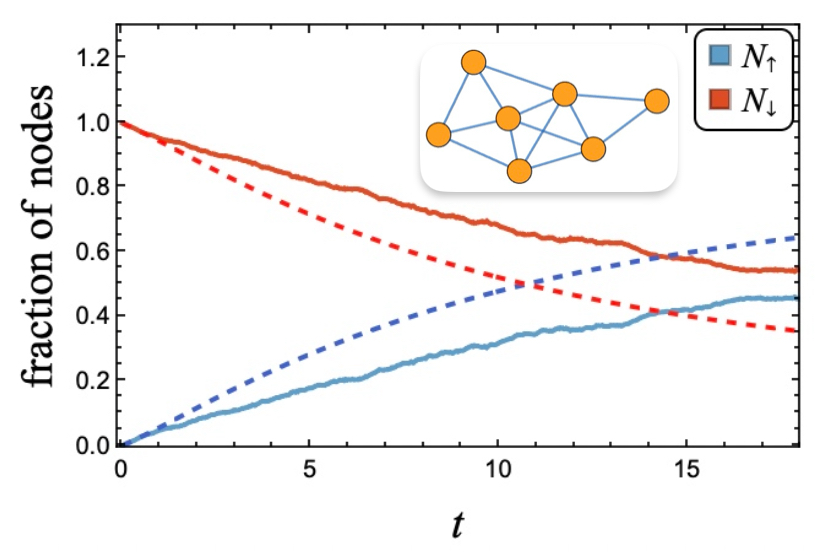}}
\hfill
\subfloat[]{\includegraphics[width=.5\linewidth]{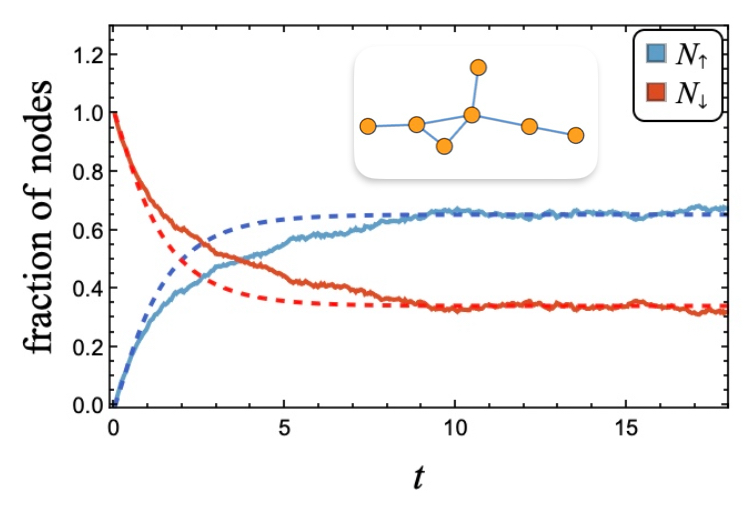}}
    \caption{The fraction of nodes $\langle N_{\uparrow,\downarrow}\rangle_t/N$ in four different network topologies, (a) $\bar k=2.50$ and $\overline{\Delta k^2}=0.333$ (b) $\bar k=2.00$ and $\overline{\Delta k^2}=0.667$ (c) $\bar k=3.71$ and $\overline{\Delta k^2}=2.571$ (d) $\bar k=2.00$ and $\overline{\Delta k^2}=1.333$. Blue curve denotes the fraction of spin-up nodes $N_{\uparrow,\downarrow}$ and red curve is for the spin down nodes. The dashed represents the mean field calculation by choosing $z=\bar{k}$}
    \label{fig:Nud}
\end{figure*}

\begin{figure*}
    \centering
\subfloat[]{\includegraphics[width=.5\linewidth]{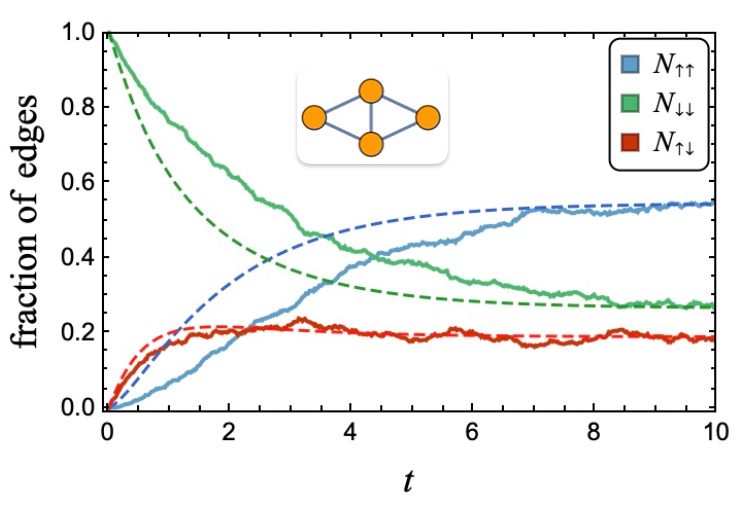}}
\hfill
\subfloat[]{\includegraphics[width=.5\linewidth]{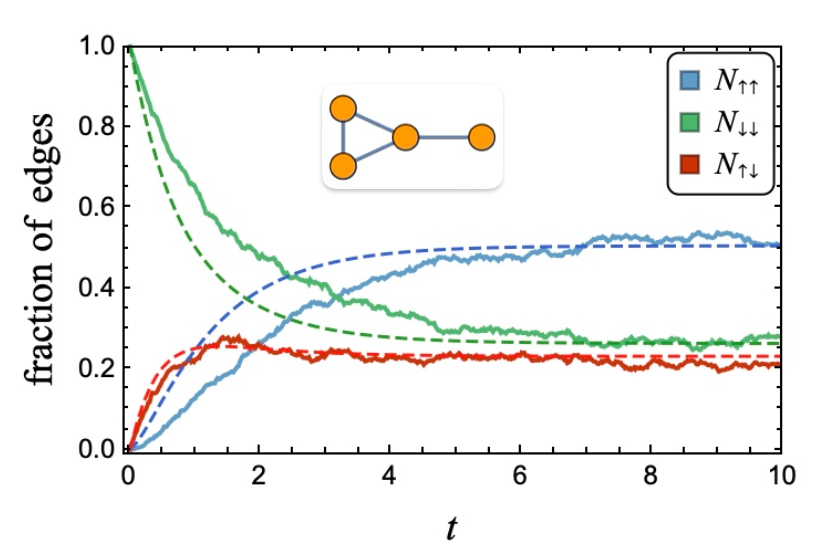}}
\hfill
\subfloat[]{\includegraphics[width=.5\linewidth]{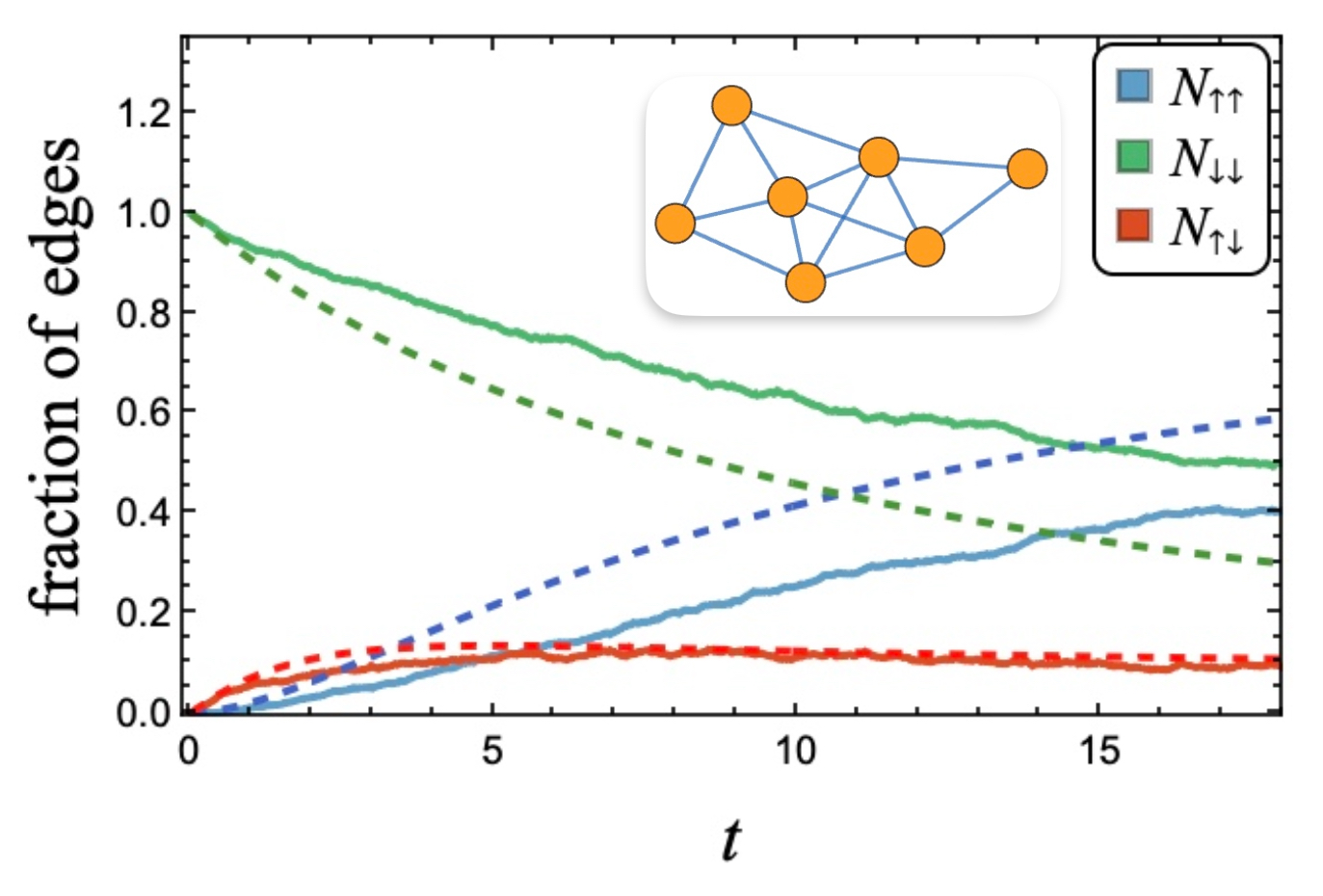}}
\hfill
\subfloat[]{\includegraphics[width=.5\linewidth]{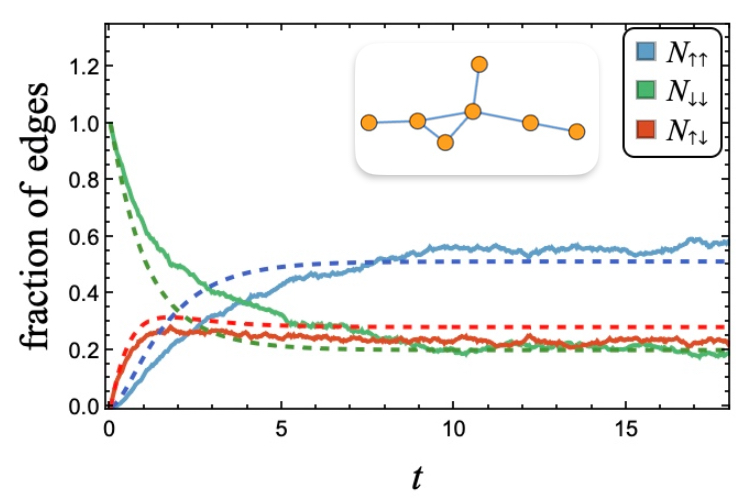}}
    \caption{The fraction of edges $\langle N_{\uparrow\uparrow,\uparrow\downarrow,\downarrow\downarrow}\rangle_t/N$ in four different network topologies, (a) $\bar k=2.50$ and $\overline{\Delta k^2}=0.333$ (b) $\bar k=2.00$ and $\overline{\Delta k^2}=0.667$ (c) $\bar k=3.71$ and $\overline{\Delta k^2}=2.571$ (d) $\bar k=2.00$ and $\overline{\Delta k^2}=1.333$. Blue curve denotes the fraction of up-up edges $N_{\uparrow\uparrow}$, green curve is for the down-down edges, and blue curve is for $N_{\uparrow\downarrow}$ the down-down edges. The dashed represents the mean field calculation by choosing $z=\bar{k}$.}
    \label{fig:Nuudd}
\end{figure*}

Therefore, any physical quantity diagonalized in spin eigenbasis related to $\sigma$ can be directly expressed by 
\begin{equation}
    O(\sigma)\rightarrow O_{N_\uparrow}
\end{equation}
This approximation largely reduce the degrees of freedom needed for the computation, especially for large $N$ network. 
This approach also shows that network topology can effectively represent the spin-lattice hypercube with fractional dimensionality under the mean-field framework. \textcolor{black}{In Fig.~\ref{fig:mean_spin}, we demonstrate the validity of the mean-field approximation. In Fig.~\ref{1}, the local magnetization trajectories of all nodes remain consistent with the global trajectory, reflecting the validity of the mean-field approximation. However, in Fig.~\ref{2}, a constant deviation in the trajectories of all nodes is observed due to a larger degree disparity, except for the trajectory of the node (indigo color) which is center of the network. In Fig.~\ref{3}, a wider band of the trajectories of each nodes is also observed with a lower degree density. Those observations show that the breakdown of the mean-field approximation happens in the presence of significant degree heterogeneity and sparse network. Therefore, we expect the mean-field analysis only works in a fairly dense homogeneous network where the coherence is maintained significantly longer.}


\subsection{Open quantum dynamics in mean field}



With mean field approximation, all spin configuration with the same $N_\uparrow$ would be degenerate. 
The reduced population (diagonal component in density matrix) by summing over all the degenerate configurations $\sigma$ reads

\begin{equation}
    P_{N_\uparrow}=\sum_{\sigma\in \left\{\sigma\Big| \substack{E(\sigma)=E_{N_\uparrow} \\ m(\sigma)=2N_\uparrow-N}\right\}}\rho_{\sigma\sigma}
\end{equation}

The quantum master equation for population dynamics becomes

\begin{equation}
\begin{aligned}
\dot{P}_{N_\uparrow} =& (N_\downarrow + 1) W_{N_\uparrow, N_\uparrow - 1} P_{N_\uparrow - 1} 
\\
&+ (N_\uparrow + 1) W_{N_\uparrow, N_\uparrow + 1} P_{N_\uparrow + 1}\\
&- (N_\uparrow W_{N_\uparrow - 1, N_\uparrow} +  N_\downarrow W_{N_\uparrow + 1, N_\uparrow}) P_{N_\uparrow}
\end{aligned}
\end{equation}
where the reduced transfer rate is defined as
\begin{equation}
\begin{aligned}
W_{N_\uparrow+1, N_\uparrow} =& \gamma
     \left(\omega_{N_\uparrow + 1,N_\uparrow}\right)
\end{aligned}
\end{equation}
with 
\begin{equation}
    \omega_{N_\uparrow + 1,N_\uparrow}=E_{N_\uparrow + 1} - E_{N_\uparrow}
\end{equation}
and the time evolution of any physical observables therefore can be evaluated by
\begin{equation}
\langle O\rangle_t=\sum_{N_\uparrow}P_{N_\uparrow}(t)O_{N_\uparrow}
\end{equation}


In both Fig.~\ref{fig:Nud} and \ref{fig:Nuudd}, they show that the mean field method is only controlled by the effective dimensionality $z$ and well captures the features while approximating a large amount of degrees of freedom. In Fig.~\ref{fig:Nud}, we compare the time evolution of the fraction of nodes $N_{\uparrow,\downarrow}$ to the mean field calculation.  Under the presence of a weak external field $h$, the number of spin-up nodes eventually exceeds the number of spin-down nodes. The final ratio in thermal equilibrium between two type of nodes are determined by the temperature $\beta$ and the strength of the field $h$. The results show that the networks with higher density and lower degree disparity exhibits greater rigidity, and the spin inversion process takes significantly longer. This structural dependence of relaxation dynamics parallels observations in social networks, where denser connectivity and reduced heterogeneity create more rigid structures that resist rapid state changes and exhibit prolonged consensus formation times \cite{lee2019social}.

In Fig.~\ref{fig:Nuudd}, we compare the time evolution of the fraction of the edges  $N_{\uparrow\uparrow,\uparrow\downarrow,\downarrow\downarrow}$ to the mean field calculation. The number of edges connecting spin-up and spin-down nodes should be optimally minimized. However, due to the constraints imposed by the network topology, some connections between opposite-spin nodes are unavoidable, representing a trade-off inherent to the structure of the network. As a result, the number of edges between opposite-spin nodes saturates earlier than other types of edges. This scenario may change significantly when rewiring effects are introduced, allowing nodes to disconnect and reconnect based on their spin states. Such adaptive dynamics can fundamentally alter the evolution of the network and the resulting spin configurations.

\section{Conclusion}
\label{sec:V}

\textcolor{black}{Our study quantitatively demonstrates that network topologies such as mean degree $\bar k$ and degree disparity $\overline{\Delta k^2}$ can alter decoherence time in a networked quantum system, establishing an illustrative framework where open quantum dynamical processes in a network system can be engineered through topological design. Although this study is restricted to small networks, it is plausible that the sensitivity of the decoherence time to network density and degree disparity could persist in larger-scale networks in a similar way. 
We acknowledge that our conclusions are based on small networks only. Verifying these trends in networks with 50 or more nodes is essential future work. Finite size effects may alter the relationship between topology and coherence time in ways we cannot predict from our current data.} Similar discussion on the quantum coherence in other topological quantum systems can be found in~\cite{nie2020topologically,bahri2015localization,yao2021topological}. Nevertheless, the connection between network topology and quantum coherence provides insights into how complex systems might naturally develop structures that preserve quantum coherence, potentially explaining observed quantum transport phenomena in complex network systems \cite{greentree2004coherent,schreiber104silberhorn,scholes2013lessons,Nakamura2021-bh}. 
\textcolor{black}{In classical network, studies of cooperation evolution on adaptive networks demonstrate how local interactions can shape global network structure \cite{gross2008adaptive, sayama2013modeling, lee2018evolutionary, lee2025enhancing, lee2025granular}, whereas analogous studies in quantum networks remain largely unexplored. 
We expect that our quantitative framework may serve as a pioneering step toward future studies of quantum game theory on networked systems.}

The mean-field framework we developed addresses the scalability challenge in quantum network simulation. The scalability is essential for understanding quantum transport and propagation in real-world-sized network system. \textcolor{black}{Although our mean-field approximation provides a simple prescription extending the system dynamics to a large-sized static homogeneous Ising network, 
the current mean-field framework does not capture heterogeneity effects such as degree disparity and fails at sparse network. Extending the mean-field approach to include additional topological parameters such as network density and degree disparity remains an open question. Nevertheless, for large homogeneous networks where exact simulation is computationally prohibitive, this mean field approach provides a tractable alternative.}   

\textcolor{black}{While the physical trends identified in this work are intuitive and largely expected, our work presents a quantitative and illustrative framework, providing a systematic and transparent way to connect network structure, especially average degree and degree disparity, to decoherence dynamics in an open quantum Ising network.
Although future developments using scalable quantum learning algorithms~\cite{valenti2022scalable} and tensor-network methods~\cite{luchnikov2019simulation} may extend our current work to larger network sizes, the conclusions of this work remain speculative but offer encouraging indications.}
\appendix

\section{Mean field approximation}
\label{app:kud}

Here we present the details of the relation between average spin and average degree in mean field approximation. Eq.~\eqref{eq:mean_cond} claims the mean degree of the network $\bar{k}$ is equal to the mean degree of the spin-up nodes
$\bar{k}_\uparrow$ as well as the spin-down nodes $\bar{k}_\downarrow$. This can also be algebraically shown by starting from the identity
\begin{equation}
\begin{aligned}
\label{eq:mean}
    &\sum_i k_i \bar{s}_i=\sum_{i}k_i\sigma_i=\sum_{i}k_i\left[\theta(\sigma_i)-\theta(-\sigma_i)\right]\\
    &=N_\uparrow\bar{k}_\uparrow-N_\downarrow\bar{k}_\downarrow
\end{aligned}
\end{equation}
where $k_i$ is the degree of node $i$, $\sigma_i \in \{+1,-1\}$ is the spin eigenvalues of each spin state of the node $i$. If we impose the mean-field condition that 
\begin{equation}
    \bar{s}_i(\sigma)=\frac{m(\sigma)}N=\frac{N_\uparrow-N_\downarrow}N
\end{equation}
where $m(\sigma)$ is the total magnetization of the spin configuration $\sigma$, then the identity in \eqref{eq:mean} becomes 

\begin{equation}
    \bar{k} (N_\uparrow-N_\downarrow)=N_\uparrow\bar{k}_\uparrow-N_\downarrow\bar{k}_\downarrow
\end{equation}

Together with the property of the mean degrees, 
\begin{equation}
    N\bar k=N_\uparrow\bar{k}_\uparrow+N_\downarrow\bar{k}_\downarrow
\end{equation}
eventually we can conclude that 
\begin{equation}
    \bar{k}=\bar{k}_\uparrow=\bar{k}_\downarrow
\end{equation}

\section{number of pairs under mean field approximation}
\label{app:mf}
\textcolor{black}{With the definitions in \eqref{eq:n} and \eqref{eq:nn}, we can derive}
\begin{equation}
\begin{aligned}
    &\sum_ik_i\theta(\sigma_i)=2N_{\uparrow\uparrow}+N_{\uparrow\downarrow}=N_\uparrow \bar{k}_\uparrow\\
    &\sum_ik_i\theta(-\sigma_i)=2N_{\downarrow\downarrow}+N_{\uparrow\downarrow}=N_\downarrow \bar{k}_\uparrow
\end{aligned}
\end{equation}
\textcolor{black}{and similarly we have}

\begin{equation}
\begin{aligned}
    &\sum_ik_i\bar{s}_i\theta(\sigma_i)=2N_{\uparrow\uparrow}-N_{\uparrow\downarrow}\\
    &\sum_ik_i\bar{s}_i\theta(-\sigma_i)=2N_{\downarrow\downarrow}-N_{\uparrow\downarrow}
\end{aligned}
\end{equation}

\textcolor{black}{Using \eqref{eq:mfc} and \eqref{eq:mean_cond}, the above equations become}

\begin{equation}
\begin{aligned}
    &2N_{\uparrow\uparrow}+N_{\uparrow\downarrow}=zN_\uparrow \\
    &2N_{\downarrow\downarrow}+N_{\uparrow\downarrow}=zN_\downarrow 
\end{aligned}
\end{equation}
\textcolor{black}{and}
\begin{equation}
\begin{aligned}
    &2N_{\uparrow\uparrow}-N_{\uparrow\downarrow}=zN_\uparrow\frac{N_\uparrow-N_\downarrow}{N}\\
    &2N_{\downarrow\downarrow}-N_{\uparrow\downarrow}=zN_\downarrow\frac{N_\uparrow-N_\downarrow}{N}
\end{aligned}
\end{equation}

\textcolor{black}{Imposing the constraint in \eqref{eq:L} and solving the resulting four equations yields the mean-field representations of $N_{\uparrow\uparrow}$, $N_{\uparrow\downarrow}$, and $N_{\downarrow\downarrow}$ in \eqref{eq:mf_nn}.}
\bibliographystyle{apsrev4-2}
\bibliography{ref}


\end{document}